\documentclass[numberedappendix,apj,twocolumn]{emulateapj}
\usepackage[colorlinks=true,citecolor=blue,linkcolor=blue]{hyperref}

\def\ltsima{$\; \buildrel < \over \sim \;$}

\def\lsim{\lower.5ex\hbox{\ltsima}}
\def\gtsima{$\; \buildrel > \over \sim \;$}
\def\gsim{\lower.5ex\hbox{\gtsima}}

\begin{document}

\title{Extrapolating the Evolution of Galaxy Sizes to the Epoch of
Reionization}

\author{J. Stuart B. Wyithe\altaffilmark{1}, Abraham Loeb\altaffilmark{2}}

\email{swyithe@unimelb.edu.au; aloeb@cfa.harvard.edu}

\altaffiltext{1}{School of Physics, University of Melbourne, Parkville, Victoria 3010, Australia}

\altaffiltext{2}{Astronomy Department, Harvard University, 60 Garden
Street, Cambridge, MA 02138, USA}

\begin{abstract}

We use data on the high-redshift evolution of the size distribution
and luminosity function of galaxies to constrain the relationship
between their star formation efficiency and starburst lifetime. Based
on the derived scaling relations, we predict the angular sizes and
average surface brightnesses of faint galaxies that will be discovered
with {\em JWST}. We find that {\em JWST} will be able to resolve
galaxies at the magnitude limit $m_{\rm AB}<31$ out to a redshift 
of $z\sim 14$. The next generation of large ground-based telescopes will resolve all
galaxies discovered with {\em JWST},  provided they are sufficiently
clumpy to enable detection above the bright thermal sky. We combine our constraints with simple models for self
regulation of star formation, and show that feedback from supernovae
at redshifts $z\ga 3$ is likely mediated through momentum transfer,
with the starburst timescale set by the lifetime of the massive stars
rather than the dynamical time in the host galactic disk.

\end{abstract}


\section{Introduction}

The luminosity function of Lyman-break galaxy candidates discovered at
$z\ga6$ in the Hubble Ultra-Deep Field is described by a Schechter
function with a characteristic luminosity that decreases towards
higher redshift as expected from the dark matter halo mass-function
(e.g. Munoz \& Loeb 2010). At low luminosities, the luminosity
function is fit as a power-law with a faint end slope of
$\alpha\approx -1.8$ that is roughly independent of redshift (McLure
et al.~2009; Bouwens et al. 2010; Yan et al. 2010). The shape and
overall density of the luminosity function is the primary observable
that must be reproduced by any successful model of galaxy formation
(e.g. Rai{\v c}evi{\'c}, Theuns \& Lacey 2010, Salvaterra, Ferrara \&
Dayal 2010; Trenti et al. 2010).

In a recent study, Oesch et al.~(2009) have measured the redshift
evolution of the scale-length of galactic disks (see also Ferguson et
al.~2004; Bouwens et al. 2004). Among a sample of galaxies with
constant luminosity, the half-light galaxy radius was found to scale
as,
\begin{equation}
R_{\rm gal} \propto (1+z)^{-m},
\end{equation}
with $m = 1.12 \pm 0.17$. This evolution is consistent with the
inverse relation between virial radius and redshift [$R_{\rm gal}
\propto (1+z)^{-1}$], that is expected for dark-matter halos assuming
a constant halo mass to luminosity ratio (Ferguson et al.~2004; Oesch
et al.~2009).

In this {\it Letter} we model the evolution of galactic disk radii for
different parameterised models of self-regulated star formation. We
parameterise our models in such a way that the constraints based on
the observed size of galaxies are orthogonal to those derived from the
slope of the galaxy luminosity function. Based on an empirical
determination of parameters of this model, we predict the expected
size distribution of galaxies in future surveys with {\em JWST}, and
show that existing observations already limit the physical processes
that govern the global properties of star formation at high redshifts.
Where required, we adopt the standard set of cosmological parameters
(Komatsu et al. 2009), with density parameters $\Omega_{\rm b}=0.044$,
$\Omega_{\rm m}=0.24$ and $\Omega_{\Lambda}=0.76$ for the matter,
baryon, and dark energy fractional density, respectively, and
$h=0.72$, for the dimensionless Hubble constant.

\section{Model}

\begin{figure*}[hptb]
\begin{center}
\vspace{3mm}
\includegraphics[width=18.cm]{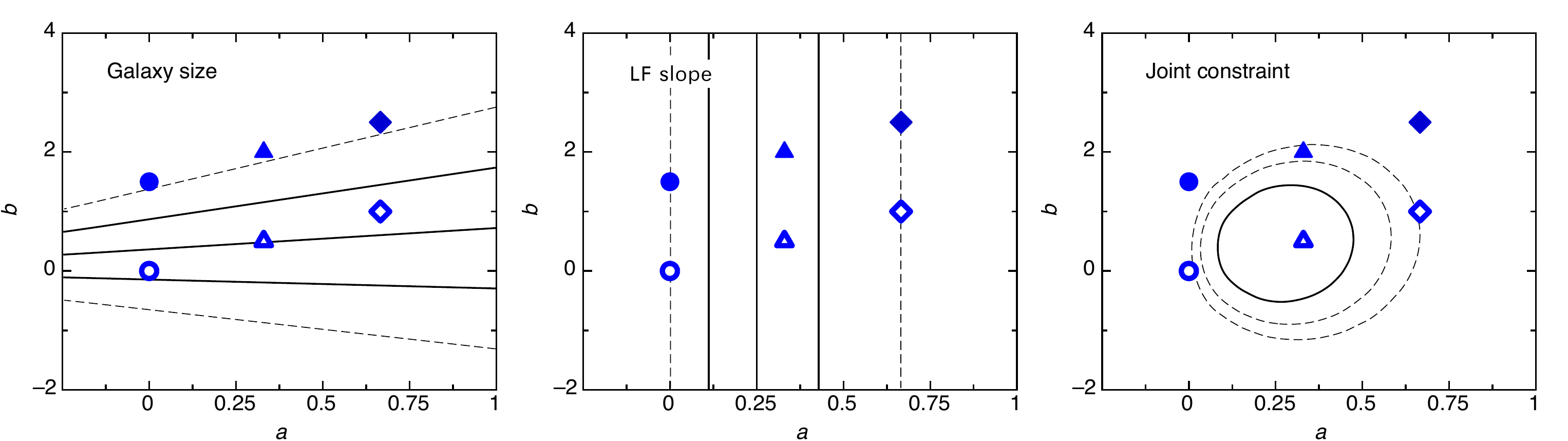}
\caption{\label{fig1} {\em Left Panel:} Contours of parameter
combinations $(a,b)$ that give the best fit observed value $m=1.12$
(central solid line), as well as the $\pm 1$-sigma (outer solid lines)
and $\pm2$-sigma combinations (dashed lines). {\em Central Panel:}
Contours of parameter combinations $(a,b)$ that give the best fit
observed value $\alpha^\prime=0.8$ (central solid line), as well as
the $\pm 1$-sigma (outer solid lines) and $\pm2$-sigma (dashed lines)
combinations. {\em Right Panel:} The combined constraint from the
observables $m$ and $\alpha$, derived from the product of likelihoods
$\mathcal{L}_{m}=\exp[-(m-1.12)^2/2(0.17)^2]$ and
$\mathcal{L}_{\alpha}=\exp[-(\alpha^\prime-0.8)^2/2(0.08)^2]$. The
contours represent the 32\%, 11\% and 4.5\% likelihoods, which for a
Gaussian distribution contain 68\%, 90\% and 95.4\% of the
probability, corresponding to the 1-sigma, 1.5-sigma and 2-sigma
ranges for parameter pairs. The model assumes $g=0$. For comparison,
the solid and open circles show the example where there is no
feedback, but the starburst timescale is proportional to the dynamical
time of the galaxy ($a=0$, $b=3/2$) or the average lifetime of massive
stars ($a=0$, $b=0$), respectively.  Similarly, the solid and open
diamond points show the example of the model in Wyithe \& Loeb~(2003)
that proposes star-formation is limited by the production of the
binding energy of the galactic gas in the form of SN-driven winds with
$t_{\rm SB}>t_\star$ ($a=2/3$, $b=5/2$) and $t_{\rm SB}<t_\star$
($a=2/3$, $b=1$). Finally, the solid and open triangle points show the
example of a model where star formation is limited by SN-driven winds
which deposit momentum rather than energy into the galactic gas, with
$t_{\rm SB}>t_\star$ ($a=1/3$, $b=2$) and $t_{\rm SB}<t_\star$
($a=1/3$, $b=1/2$).  }
\end{center}
\end{figure*}

To model high redshift galaxy properties we begin with an expression
for the UV luminosity of a galaxy within a dark matter halo of mass
$M_{\rm halo}$,
\begin{equation}
\label{luminosity}
L \propto f_\star M_{\rm halo} \frac{\min(t_{\rm SB},t_\star) }{t_{\rm SB}\,t_\star}, 
\label{eq:L}
\end{equation}
where $f_\star$ is the fraction of baryons turned into stars, $t_{\rm
SB}$ is the lifetime of the starburst, and $t_\star$ is the average
lifetime of the massive stars ($\ga 8M_\odot$) that affect the
feedback through supernova (SN) explosions and dominate the UV
luminosity output $L$.  The lifetime of main-sequence stars, $t_{\rm
ms}=10^{10}{\rm yr} (M_{\rm star}/M_\odot)^{-2.5}$ (Hansen, Kawaler \&
Steven 1994), implies an average delay time for supernova feedback of
$t_\star\sim 7\times10^6$ yr for an initial mass function of massive
stars $dN/dM_{\rm star}\propto M_{\rm star}^{-2.35}$ (Scalo~1986). The
value of $t_{SB}$ is thought to be related to the dynamical time
($t_{\rm dyn}$) of the galactic disk (Kennicutt~1998), which scales as
the age of the Universe $t_H$, namely $t_{\rm dyn}\sim 3\times 10^{-3}
t_H \approx 3\times 10^6~{\rm yr}[(1+z)/7]^{-3/2}$.  Equation
(\ref{eq:L}) implies that the luminosity could become independent of
the starburst lifetime at sufficiently high redshifts $z\ga 3$ for
which $t_{\rm dyn}<t_\star$. The mass-to-light ratio is governed by
$f_\star$ and $t_{\rm SB}$, each of which can depend on both $M_{\rm
halo}$ and $z$. We therefore parameterise a general class of models
for high redshift galaxy formation using the ratio,
\begin{equation}
\label{parameter}
f_\star \frac{\min(t_{\rm SB},t_\star) }{t_{\rm SB}\,t_\star}\propto
M_{\rm halo}^a (1+z)^b.
\end{equation}
Physically, this quantity is proportional to the inverse of the
mass-to-light ratio. This parameterisation can be used to describe a
range of physical models that predict the unknown variation of the
star-formation efficiency and starburst lifetime with redshift and
halo mass. Each particular model of the self-regulation of high
redshift star formation will yield distinct values for the parameters
$a$ and $b$.

In order to compare the resulting grid of models with the observed
evolution in galaxy size we require two further ingredients, namely
the virial relation (simplified form valid at high redshift),
\begin{equation}
\label{virial}
M_{\rm halo}\propto R_{\rm vir}^3 (1+z)^3,
\end{equation}
and the parameterisation,
\begin{equation}
\label{gal}
R_{\rm gal} \propto R_{\rm vir} (1+z)^g ,
\end{equation}
to describe the relation between virial radius $R_{\rm vir}$ and the
galactic disk scale radius $R_{\rm gal}$. In the latter expression we
expect a parameterisation with $g=0$ if the disk size\footnote{If the
gas disk becomes stable to fragmentation at a radius beyond which
there is a significant fraction of gas by mass, then the half mass
radius of the stellar disk may not equal the scale radius of the
gas-disk. However, we find that the disk becomes stable (based on
Toomre Q criterion) only at 3-4 scale radii, and therefore that there
is a very small fraction of mass at these large radii. Thus, we adopt
the assumption $g=0$ throughout our analysis.} is set by the adiabatic
model of Mo, Mao \& White~(1998). The above set of relations can be
used to find the predicted evolution of the galaxy radius with
redshift at a fixed luminosity, yielding
\begin{equation}
\label{m}
m = 1 -g + \frac{1}{3}\frac{b}{a+1},
\end{equation}
which can be compared with the observed value of $m=1.12\pm0.17$
(Oesch et al. 2009).

It is also possible to use the observed slope of the luminosity
function to constrain the model, thus breaking the degeneracy between
the parameters $a$ and $b$ that arises from application of
equation~(\ref{m}). Modelling the galaxy luminosity function using the
halo mass function ($dn/dM_{\rm halo}$) and the simple parameterised
model described in equations~(\ref{luminosity}-\ref{parameter}) above
we find
\begin{eqnarray}
\label{alpha}
\nonumber \alpha &=& \frac{d\log{n}}{d\log{L}}=
\frac{d\log{n}}{d\log{M_{\rm halo}}}\left|\frac{d\log{M_{\rm
halo}}}{d\log{L}}\right| \\ &=&
\frac{1}{a+1}\frac{d\log{n}}{d\log{M_{\rm halo}}}.
\end{eqnarray}
In the mass range $10^8M_\odot<M_{\rm halo}<10^{10.5}M_\odot$ (near or
below the non-linear mass scale at $z\sim6$), the mass function has
the local power-law slope $-2.05\ga{d\log{n}}/{d\log{M_{\rm halo}}}
\ga-2.45 $.  At low luminosities the high redshift galaxy luminosity
function is fit as a power-law with faint end slope $\alpha=-1.8$
(McLure et al.~2009; Bouwens et al. 2010; Yan et al. 2010). While the
uncertainty in $\alpha$ is large at $z\gsim7$, it is well constrained
in the redshift range $4\la z\la6$. We estimate a 10\% uncertainty in
the value of $({d\log{n}}/{d\log{M_{\rm halo}}})$, dominated by the
uncertain mass of the host halos. We therefore define
$\alpha^\prime\equiv \alpha \times ({d\log{n}}/{d\log{M_{\rm
halo}}})^{-1}$, and adopt the constraint $\alpha^\prime= 1/(a+1) =
0.8\pm0.1$ based on equation~(\ref{alpha}).

\subsection{Parameter constraints}

\begin{figure*}[hptb]
\begin{center}
\includegraphics[width=18.cm]{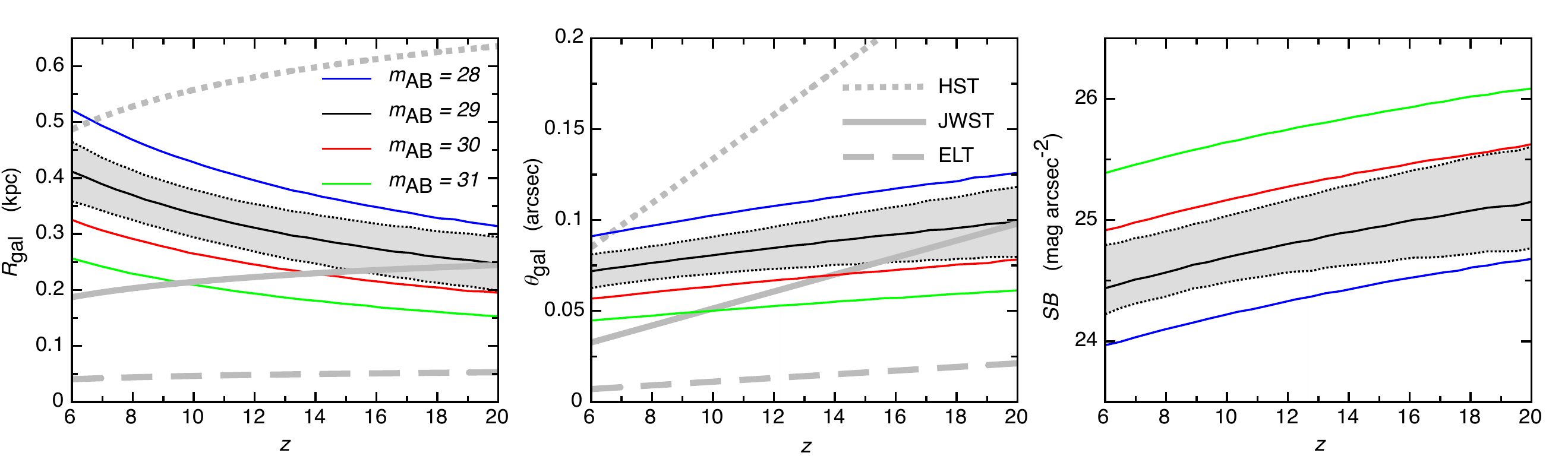}
\caption{\label{fig2} The projected relation between galaxy size and
redshift, plotted for four values of apparent magnitude. The {\em
Left} and {\em Central} panels show the physical ($R_{\rm gal}$) and
apparent angular sizes ($\theta_{\rm gal}$) respectively. For
comparison (thick grey lines), we also plot the indicative angular
resolution $\Delta\theta$ of telescopes with diameters corresponding
to {\em HST} ($D_{\rm tel}=2.5$\,m), {\em JWST} ($D_{\rm tel}=6.5$\,m)
and ELT ($D_{\rm tel}=30$\,m). The {\em Right} panel shows the
relation between surface brightness (averaged within the scale radius)
as a function of redshift. In each panel, the grey band around the
case of $m_{\rm AB}=29$\,mag shows the $68\%$ range of uncertainty on
the mean.  }
\end{center}
\end{figure*}

The left hand panel of Figure~\ref{fig1} shows contours of parameter
combinations $(a,b)$ that give the best fit observed value $m=1.12$,
as well as the $\pm 1$-sigma and $\pm2$-sigma combinations
(equation~\ref{m} assuming $g=0$). The central panel of
Figure~\ref{fig1} shows contours of parameter combinations $(a,b)$
that give the best fit observed value $\alpha^\prime=0.8$, as well as
the $\pm 1$-sigma and $\pm2$-sigma combinations (again assuming
$g=0$). The right hand panel shows the combined constraint from both
observables, with the contours representing the 32\%, 11\% and 4.5\%
likelihoods, which for a Gaussian distribution correspond the 1-sigma,
1.5-sigma and 2-sigma regions for parameters combinations.  We find
$a=0.3\pm0.15$ and $b=0.5\pm0.6$ (68\% range on individual parameters).

\section{The size of high redshift galaxies}

Our empirical constraints can be used to extrapolate galaxy size out
to the higher redshifts and lower luminosities that will be observed
by the next generation of telescopes. Beginning with
equations~(\ref{luminosity}-\ref{parameter}) we get,
\begin{equation}
L\propto M_{\rm halo}^{a+1}(1+z)^b,
\end{equation}  
which when combined with equations~(\ref{virial}-\ref{gal}) implies, 
\begin{eqnarray}
\nonumber R_{\rm gal}&\propto&
L^{\frac{1}{3(1+a)}}(1+z)^{-(1-g+\frac{b}{3(1+a)})} \propto
L^{\frac{1}{3(1+a)}}(1+z)^{-m}\\ &=& R_0 \left(\frac{D_{\rm
L}(z)}{D_{\rm L}(z_0)}\right)^\frac{2}{3(1+a)} 10^{\frac{m_{\rm
AB,0}-m_{\rm AB}}{7.5(a+1)}}\left(\frac{1+z}{1+z_0}\right)^{-m},
\end{eqnarray}
where $D_{\rm L}$ is the luminosity distance, $m_{\rm AB}$ is the
apparent galaxy magnitude, and $m_{\rm AB,0}$ is the apparent
magnitude of galaxies from which the normalization of the relation
($R_0$) is measured at redshift $z_0$. The normalisation of this
relation is calibrated using observed galaxy sizes at $z\sim7-8$
(Oesch et al.~2009).  We average the prediction over the three
independent points\footnote{Uncertainty in $R_{\rm 0}$ is the
standard error on the mean.} having $(z_0,R_{0},m_{\rm AB,0}) =
(7,0.5\pm0.1,27.8)$, $ (7,0.75\pm0.1,26.6)$, and $(8,0.4\pm0.1,28.1)$,
and over distributions $m=1.12\pm0.17$ and $\alpha=0.3\pm0.15$. The
resulting relation is shown in Figure~\ref{fig2} as a function of
redshift for four values of apparent magnitude. In the case of $m_{\rm
AB}=29$\,mag, the grey band shows the $68\%$ range of uncertainty on
the mean radius.  The left and central panels show the physical and
apparent angular sizes ($\theta_{\rm gal}$), respectively. Galaxies of
fixed apparent magnitude have smaller physical sizes (but larger
angular sizes) at higher redshift.

For comparison, we also plot in Figure~\ref{fig2}, the indicative
angular resolutions of telescopes with diameters corresponding to {\em
HST} ($D_{\rm tel}=2.5$\,m), {\em JWST} ($D_{\rm tel}=6.5$\,m) and an
extremely large telescope (ELT; $D_{\rm tel}=30$\,m)
\begin{equation}
\Delta\theta = \frac{1.22 \lambda}{D_{\rm tel}} \approx
0.085\left(\frac{1+z}{7}\right)\left(\frac{D_{\rm
tel}}{2.5}\right)^{-1}.
\end{equation}
Here we have calculated the resolution at the observed wavelength of
the redshifted Ly$\alpha$ line $\lambda = 1216 (1+z)$\AA. This simple
comparison suggests that galaxies with an apparent magnitude of
$m_{\rm AB}=28$\,mag (1 magnitude fainter than the {\em HST} WFC3/IR
limit) have an angular size of $\theta_{\rm gal}\sim0.1''$, which is
at the limit of {\em HST} resolution (Oesch et al.~2009). Fainter galaxies at
higher redshifts cannot be resolved by {\em HST}. The larger aperture
of {\em JWST} will allow the study of galaxy structure at higher
redshift and fainter fluxes (Windhorst et al.~2008). For example,
galaxies with $m_{\rm AB}=29$\,mag will be resolved out to at least
$z\ga15$. Moreover, {\em JWST} will be able to resolve galaxies at the
magnitude limits of $m_{\rm AB}<30$ and $m_{\rm AB}<31$ out to $z\sim
10$ and $z\sim14$, respectively. Galaxies at yet higher redshifts or
fainter fluxes would appear point-like. Bright galaxies are known to
become rarer toward high redshift (e.g. Bouwens et al.~2010; Yan et
al.~2010), and the discovery of such very high redshift galaxies by
{\em JWST} could be limited by its relatively small field of view
(e.g. Munoz \& Loeb~2010). However, our extrapolation suggests that an
{\em ELT} would be able to resolve all galaxies detectable with {\em
JWST}.

The above discussion refers only to angular resolution and neglects
surface brightness sensitivity. Calculation of the details of surface
brightness sensitivity (see, e.g. Windhorst et al.~2008) are beyond
the scope of this {\em Letter}. However, we note that faint,
high-redshift galaxies are always below the surface brightness of the
sky.  Therefore, in order to measure the details of a high redshift
galaxy profile, an {\em HST} observation must achieve a very high
signal-to-noise ratio on the zodiacal sky (of order $10^3$ per pixel)
so that the background would be reliably subtracted (Hathi et
al.~2008).  The right panel of Figure~\ref{fig2} shows the average
surface-brightness within the scale-radius of galaxies of various
magnitudes as a function of redshift. Galaxies with $m_{\rm
AB}=29$\,mag show $\sim 24.5$ magnitudes per square arcsecond at
$z\sim6$, becoming fainter by only a modest amount towards
$z\sim20$. The plotted curves should be compared to the space based
value for the zodiacal sky in the K-band of $\sim 22.5$ magnitudes per
square arcsecond (Windhorst et al.~2010), or the ground based K-band
value of $\sim 14$ magnitudes per square arcsecond for the thermal
sky.  The difference in sky brightness between the ground and space
implies that {\em HST} is equivalent in sensitivity (but not
in resolution) to a 150 m ground-based telescope for the purpose of
imaging resolved high redshift galaxies. Thus, a ground-based {\em
ELT} will not be competitive with {\em HST} or {\em JWST}, since it
must overcome the much brighter sky. As a result, even though all high
redshift galaxies discovered with {\em JWST} could be resolved by an
{\em ELT}, their surface brightness will render their extended
structure undetectable.  However, high redshift galaxies are observed
to be very clumpy owing to the presence of star forming regions (Hathi
et al.~2008), and an {\em ELT} will be more sensitive to these
unresolved clumps than to diffuse structure, owing to the higher
resolution in addition to larger collecting area. Indeed, {\em HST} is
comparable to {\em only} a 20 m ground based telescope with respect to
point source sensitivity (but without the resolution of a 20 m
aperture). Thus, if high redshift galaxies are sufficiently clumpy,
then an {\em ELT} may be able to obtain close to the theoretical
resolution shown in the central panel of Figure~\ref{fig2}.

\section{Star Formation model constraints}

A range of simple models for the process governing star formation can
be compared to the constraints on parameters $a$ and $b$. We consider
three simple models for the possible feedback. The first is a model
where there is no self regulation of star formation. The second and
third models describe the evolution in cases where the star formation
is limited by SN feedback, and the interaction between the SN driven
wind and the galactic gas conserves energy and momentum,
respectively. In each of these three cases we consider scenarios where
the lifetime of the starburst is proportional to the galaxy
dynamical time and to the lifetime of massive stars,
respectively. Altogether, we have 6 model predictions for $m$ and
$\alpha^\prime$ with which to compare our constraints.

\subsection{Models without feedback}

We begin with feedback-free models in which the star formation
efficiency $f_\star \propto M_\star/M_{\rm halo} = const$. We first
consider the case where the natural timescale for the starburst is
proportional to the galaxy dynamical time, $t_{\rm SB}\propto
t_{\rm dyn}\propto(1+z)^{-3/2}$, which implies
\begin{equation}
 f_\star\frac{\min(t_{\rm SB},t_\star) }{t_{\rm SB}\,t_\star} \propto
 \frac{f_\star}{t_{\rm dyn}} \propto \frac{M_{\rm halo}^{0}
 (1+z)^0}{(1+z)^{-3/2}} \propto M_{\rm halo}^{0} (1+z)^{3/2},
\end{equation}
yielding $a=0$ and $b=3/2$. This parameter combination is shown by the
solid circles in Figure~\ref{fig1} for comparison with the present
constraints. We find that this simple model is inconsistent with both
the observed evolution in $R_{\rm gal}$ and the observed luminosity
function slope at the $2$-sigma level. The combined constraint rules
out this model at high confidence.

However, at the high redshifts when the dynamical time is shorter than
the average lifetime of massive stars, the same luminosity may be
achieved with a lower star formation rate than in the model described
above,
\begin{equation}
f_\star \frac{\min(t_{\rm SB},t_\star) }{t_{\rm SB}\,t_\star}\propto
\frac{f_\star}{t_\star} \propto M_{\rm halo}^{0} (1+z)^0,
\end{equation}
yielding $a=b=0$ for galaxies in this case. Such a model (shown by the
open circular symbols) is ruled out by the slope of the luminosity
function at the 2-sigma level, but is consistent with the observed
evolution in galaxy radius. The combined constraint rules out this
model at high confidence.
 
 \subsection{Models with feedback through energy conservation}

We next consider models including self-regulation of star formation by
supernovae feedback. The model of Wyithe \& Loeb~(2003) assumes that
star formation is limited by the transfer of energy from SN-driven
winds when it is equal to the binding energy to the galactic gas
(Dekel \& Woo 2002). The model asserts that stars form with an
efficiency $f_\star$ out of the gas that collapses and cools within a
dark matter halo and that a fraction $F_{\rm SN}$ of each supernova
energy output, $E_{\rm SN}$, heats the galactic gas mechanically. The
mechanical feedback will halt the star formation once the cumulative
energy returned to the gas by supernovae equals its binding energy
(assuming negligible radiative losses for a sufficiently rapid
starburst).  Hence, in this model the limiting stellar mass is set by
the condition
\begin{equation}
\label{feedback}
\frac{M_\star}{w_{\rm SN}}E_{\rm SN}F_{\rm SN} = E_{\rm b} =
\frac{1}{2}\frac{\Omega_{\rm b}}{\Omega_{\rm m}}M_{\rm halo}v_{\rm
c}^2,
\end{equation}
where $w_{\rm SN}$ is the mass in stars (in $M_\odot$) per supernova
explosion. Equations~(\ref{virial}) and (\ref{feedback}) imply that
the total mass in stars, $M_\star = (f_\star\Omega_{\rm b}/\Omega_{\rm
m})M_{\rm halo}$, scales as
\begin{equation}
M_\star \propto M_{\rm halo}^{5/3}(1+z),
\end{equation}
and hence the star formation efficiency scales as $f_\star \propto
M_{\rm halo}^{2/3}(1+z)$. Thus, smaller galaxies are less efficient at
forming stars, but a galaxy of fixed mass is more efficient at forming
stars at higher redshift.

In a model where the starburst lifetime is proportional to the galaxy dynamical
time, we find
\begin{equation}
\frac{f_\star}{t_{\rm dyn}} \propto \frac{M_{\rm halo}^{2/3}
(1+z)}{(1+z)^{-3/2}}\propto M_{\rm halo}^{2/3} (1+z)^{5/2}.
\end{equation}
This model can therefore be parameterised in terms of the combination
$a=2/3$, and $b=5/2$ for galaxies, which is shown by the solid
diamonds in Figure~\ref{fig1}. This simple supernovae driven feedback
model is inconsistent with the observed evolution in $R_{\rm gal}$ and
the luminosity function slope at the $2$-sigma level. The combined
constraint rules out this model at high confidence.

However, at high redshifts $z\ga 3$ when the disk dynamical time is
shorter than the lifetime of a massive star $t_\star\sim 10^7$yr, we
note that SN feedback will be less efficient, and the star
formation efficiency could exceed the self regulated value of
$f_\star$ described above. In particular,
\begin{equation}
\frac{f_\star}{t_\star} \propto \frac{M_{\rm halo}^{2/3} (1+z)}{t_\star} \propto M_{\rm halo}^{2/3} (1+z),
\end{equation}
yielding $a=2/3$, and $b=1$ for galaxies in this case (open diamond
symbols). Such a model is consistent with the observed evolution of
galaxy size, but inconsistent with the slope of the luminosity
function at the 2-sigma level.

\subsection{Models with feedback through momentum conservation}

Finally, we consider a model where the SN-driven winds conserve
momentum in their interaction with the galactic gas rather than
energy. In this case the limiting stellar mass is set by the condition
\begin{equation}
\label{feedback2}
\frac{M_\star}{w_{\rm SN}}{E_{\rm SN} \over c}F_{\rm SN} = \frac{1}{2}\frac{\Omega_{\rm b}}{\Omega_{\rm m}}M_{\rm halo}v_{\rm c}.
\end{equation}
In a model where the starburst
timescale is proportional to the galaxy dynamical time, we find
\begin{equation}
\frac{f_\star}{t_{\rm dyn}} \propto M_{\rm halo}^{1/3} (1+z)^{2}. 
\end{equation}
This model (solid triangles), represented by $a=1/3$ and $b=2$, is
consistent with the constraints from $\alpha$, but ruled out at the
2-sigma level by the constraints from the evolution of galaxy
radius. If instead the lifetime of massive stars sets the starburst
timescale, we find
\begin{equation}
\frac{f_\star}{t_\star} \propto M_{\rm halo}^{1/3} (1+z)^{1/2},
\end{equation}
which is represented by $a=1/3$ and $b=1/2$, and is plotted as the
open triangles in Figure~\ref{fig1}. This model is consistent with
both the constraints from $m$ and $\alpha^\prime$.

Our results imply that SN feedback in high redshift galaxies occurs
through the transfer of momentum between the SN-driven wind and the
galactic gas, and that the starburst timescale is set by the average
lifetime of massive stars rather than the dynamical time of the host
galactic disk.

\section{Discussion}

We have used the observed redshift evolution of disk sizes, and
luminosity function of galaxies to constrain the relationship between
star formation efficiency and starburst lifetime.  We find that
supernova feedback in high redshift galaxies is likely mediated
through momentum transfer with the starburst timescale dictated by the
average lifetime of the massive stars, $t_\star$. The latter result
follows naturally from the fact that the dynamical time of galactic
disks becomes shorter than $t_\star \sim 10^7$ yr at redshifts $z\ga
6$.

We extrapolated the derived scaling relations to predict the angular
sizes of galaxies at higher redshifts and fainter fluxes than
currently observed. We have found that {\em JWST} will be able to
resolve galaxies with $m_{\rm AB}<30$ and $m_{\rm AB}<31$ only out to
redshifts of $z\sim 10$ and $z\sim14$, respectively. However, if high
redshift galaxies are sufficiently clumpy, so that unresolved star
forming regions can be detected above the bright thermal sky, then the
next generation of ground-based extremely large telescopes will be
able to resolve structure in all galaxies at all redshifts detectable
by {\em JWST}.

\acknowledgements 

We thank Rogier Windhorst for helpful discussions. JSBW acknowledges
the support of the Australian Research Council.  AL was supported in
part by NSF grant AST-0907890 and NASA grants NNX08AL43G and
NNA09DB30A.

\bibliographystyle{apj}

\end{document}